# High Frequency Impedances in European XFEL

Martin Dohlus, Igor Zagorodnov[*], Olga Zagorodnova

*Deutsches Elektronen-Synchrotron (DESY), Notkestrasse 85, 22603 Hamburg, Germany*

The method of the optical approximation is used to estimate the high frequency impedances of different vacuum chamber transitions of the European XFEL beam line. The approximations of the longitudinal impedances are obtained in terms of simple one-dimensional integrals. The transverse impedances are written in analytical closed form. The analytical results are compared with the results obtained by numerical solution of Maxwell's equations.

[*] Corresponding author. Tel.:+49-040-8998-1802; fax: +49-040-8998-4305
   *E-Mail address*: igor.zagorodnov@desy.de

# I. INTRODUCTION

In the European X-ray Free Electron Laser [1] short bunches with high peak current will be transported through the beam line and the undulators. Such bunches are sources of strong short range wakefields which increase the energy spread in the bunch and spoil its emittance.

In this report we apply the method of the optical approximation [2, 3] to estimate the high frequency impedance of different vacuum chamber transitions: a symmetric rectangular laser mirror in round beam pipe, an asymmetric rectangular optical transition radiation (OTR) screen in a round beam pipe, round-to-elliptical pipe transitions and elliptical absorbers in the sections between undulators, round-to-rectangular pipe transitions in the bunch compressors. We consider the longitudinal and the transverse impedances generated by misalignment of round pipes. The analytical results are given in a closed form or as a one-dimensional integral. Almost all obtained impedances are compared to the results of direct time-domain simulations with codes ECHO [4] and CST Particle Studio [5].

Additionally, we estimate the effect of the wakefields on the beam and compare the obtained wakes to the wakes of the other elements in the beam line.

# II. OPTICAL APPROXIMATION

In recent papers [2, 3] a method to estimate the high frequency impedance of short transitions was developed.

Consider a short transition with aperture $S_{ap}$ between two pipes with apertures $S_A$ and $S_B$ as shown in Fig. 1. Let $a$ is a characteristic size of the aperture $S_{ap}$. If the bunch has a short rms length $\sigma$, $\sigma \ll a$, and the transition length $L$ between the ingoing pipe aperture $S_A$ and the outgoing pipe aperture $S_B$ is short, $L \ll a^2/\sigma$, then the high frequency longitudinal impedance is a *constant* which can be calculated by relation (Eq. (36), in [2])

$$Z_{\parallel}(\mathbf{r}_1,\mathbf{r}_2) = \frac{2\varepsilon_0}{c}\left[\int_{S_B}\nabla\varphi_B(\mathbf{r}_1,\mathbf{r})\nabla\varphi_B(\mathbf{r}_2,\mathbf{r})ds - \int_{S_{ap}}\nabla\varphi_A(\mathbf{r}_1,\mathbf{r})\nabla\varphi_B(\mathbf{r}_2,\mathbf{r})ds\right] \quad (1)$$

where $\mathbf{r}_1$ and $\mathbf{r}_2$ are offsets of the leading and the trailing particles, correspondingly, and $\varphi_A$, $\varphi_B$ are the Green's functions for the Laplacian in the ingoing and the outgoing pipe cross-sections

$$\Delta\varphi_A(\mathbf{r}_i,\mathbf{r}) = -\varepsilon_0^{-1}\delta(\mathbf{r}-\mathbf{r}_i), \; \mathbf{r}\in S_A, \quad (2)$$
$$\varphi_A(\mathbf{r}_i,\mathbf{r}) = 0, \; \mathbf{r}\in\partial S_A,$$

$$\Delta\varphi_B(\mathbf{r}_i,\mathbf{r}) = -\varepsilon_0^{-1}\delta(\mathbf{r}-\mathbf{r}_i), \; \mathbf{r}\in S_B, \quad (3)$$
$$\varphi_B(\mathbf{r}_i,\mathbf{r}) = 0, \; \mathbf{r}\in\partial S_A, \; i=1,2,$$

where $\partial S$ is the boundary of $S$, $\varepsilon_0$ is the permittivity of free space.

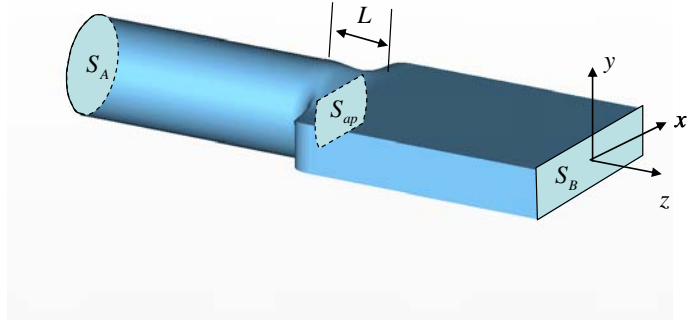

FIG 1. The geometry of transition.

With the help of Green's first identity (**n** is the outward pointing unit normal to the line element $dl$ )

$$\int_S \nabla\varphi \nabla\psi \, ds = \int_{\partial S} \varphi \partial_{\mathbf{n}}\psi \, dl - \int_S \varphi \Delta\psi \, ds, \qquad (4)$$

we can rewrite Eq. (1) (with $\varphi \equiv \varphi_B$, $\psi \equiv \varphi_A$ ) as

$$Z_\|(\mathbf{r}_1,\mathbf{r}_2) = -\frac{2\varepsilon_0}{c} \int_{\partial S_{ap}} \varphi_B(\mathbf{r}_2,\mathbf{r}) \partial_{\mathbf{n}} \varphi_A(\mathbf{r}_1,\mathbf{r}) dl, \qquad (5)$$

where we have used Eqs. (2), (3).

If we use another assignment of the functions in Eq. (4) ($\varphi \equiv \varphi_A$, $\psi \equiv \varphi_B$ ) then we obtain an alternative expression

$$Z_\|(\mathbf{r}_1,\mathbf{r}_2) = -\frac{2\varepsilon_0}{c} \int_{\partial S_{ap}} \varphi_A(\mathbf{r}_1,\mathbf{r}) \partial_{\mathbf{n}} \varphi_B(\mathbf{r}_2,\mathbf{r}) dl + \frac{2}{c}\left(\varphi_B(\mathbf{r}_1,\mathbf{r}_2) - \varphi_A(\mathbf{r}_1,\mathbf{r}_2)\right). \qquad (6)$$

To simplify the equation's notation we consider a situation when the offsets $\mathbf{r}_1$ and $\mathbf{r}_2$ lay in $x = 0$ coordinate plane. In this case we can omit the coordinate $x$ and use a new notation

$$\varphi(y,\mathbf{r}) \equiv \varphi\left(\begin{pmatrix}0\\y\end{pmatrix},\mathbf{r}\right), \quad Z_\|(\mathbf{y}) \equiv Z_\|(y_1,y_2) \equiv Z_\|\left(\begin{pmatrix}0\\y_1\end{pmatrix},\begin{pmatrix}0\\y_2\end{pmatrix}\right), \quad \mathbf{y} \equiv (y_1,y_2)^T.$$

Then the Taylor expansion for the longitudinal impedance reads

$$Z_\|(\mathbf{y}) = Z_\|(\mathbf{0}) + \mathbf{y}_{12} \nabla Z_\|(\mathbf{0}) + \frac{1}{2}(\mathbf{y}_{12}\nabla)^2 Z_\|(\mathbf{0}) + O(|\mathbf{y}_{12}|^3),$$

where the third term in the right hand side is defined through the Hessian matrix

$$(\mathbf{y}\nabla)^2 Z_\|(\mathbf{0}) \equiv \mathbf{y}^T \begin{pmatrix} \dfrac{\partial^2}{\partial y_1'^2} & \dfrac{\partial^2}{\partial y_1' \partial y_2'} \\ \dfrac{\partial^2}{\partial y_2' \partial y_1'} & \dfrac{\partial^2}{\partial y_2'^2} \end{pmatrix} Z_\|(\mathbf{y}')\bigg|_{\mathbf{y}'=0} \mathbf{y}.$$

The leading term in this expansion can be written as

$$Z_\|(\mathbf{0}) = -\frac{2\varepsilon_0}{c} \int_{\partial S_{ap}} \varphi_B(0,\mathbf{r}) \partial_{\mathbf{n}} \varphi_A(0,\mathbf{r}) dl. \qquad (7)$$

The transverse impedance is related to the longitudinal one through the Panofsky-Wenzel theorem

$$Z_y(\omega, \mathbf{y}) = \frac{c}{\omega} \frac{\partial}{\partial y_2} Z_\|(\mathbf{y}).$$

The series expansion for the transverse impedance reads

$$Z_y(\omega, \mathbf{y}) = \frac{c}{\omega}\left( \frac{\partial Z_\|}{\partial y_2}(\mathbf{0}) + y_1 \frac{\partial^2 Z_\|}{\partial y_1 \partial y_2}(\mathbf{0}) + y_2 \frac{\partial^2 Z_\|}{\partial y_2^2}(\mathbf{0}) \right) + O\left(|\mathbf{y}|^2\right)$$

It follows from Eqs. (5), (6) that the leading terms in the series expansion of the transverse impedance can be written as

$$Z_y^{(m)}(\omega) = \frac{c}{\omega}\frac{\partial Z_\|}{\partial y_2}(\mathbf{0}) = -\frac{2\varepsilon_0}{\omega} \int_{\partial S_{ap}} \varphi_B^{(d)}(\mathbf{r}) \partial_\mathbf{n} \varphi_A^{(m)}(\mathbf{r}) dl, \tag{8a}$$

$$Z_y^{(m)}(\omega) = -\frac{2\varepsilon_0}{\omega} \int_{\partial S_{ap}} \varphi_A^{(m)}(\mathbf{r}) \partial_\mathbf{n} \varphi_B^{(d)}(\mathbf{r}) dl + \frac{2}{\omega}\left(\varphi_B^{(d)}(0) - \varphi_A^{(d)}(0)\right), \tag{8b}$$

$$Z_y^{(d)}(\omega) = \frac{c}{\omega}\frac{\partial^2 Z_\|}{\partial y_1 \partial y_2}(\mathbf{0}) = -\frac{2\varepsilon_0}{\omega} \int_{\partial S_{ap}} \varphi_B^{(d)}(\mathbf{r}) \partial_\mathbf{n} \varphi_A^{(d)}(\mathbf{r}) dl, \tag{9a}$$

$$Z_y^{(d)}(\omega) = -\frac{2\varepsilon_0}{\omega} \int_{\partial S_{ap}} \varphi_A^{(d)}(\mathbf{r}) \partial_\mathbf{n} \varphi_B^{(d)}(\mathbf{r}) dl + \frac{2}{\omega}\lim_{y\to 0}\frac{\partial}{\partial y}\left(\varphi_B^{(d)}(y) - \varphi_A^{(d)}(y)\right), \tag{9b}$$

$$Z_y^{(q)}(\omega) = \frac{c}{\omega}\frac{\partial^2 Z_\|}{\partial y_2^2}(\mathbf{0}) = -\frac{2\varepsilon_0}{\omega} \int_{\partial S_{ap}} \varphi_B^{(m)}(\mathbf{r}) \partial_\mathbf{n} \varphi_A^{(q)}(\mathbf{r}) dl, \tag{10a}$$

$$Z_y^{(q)}(\omega) = -\frac{2\varepsilon_0}{\omega} \int_{\partial S_{ap}} \varphi_A^{(m)}(\mathbf{r}) \partial_\mathbf{n} \varphi_B^{(q)}(\mathbf{r}) dl + \frac{2}{\omega}\left(\varphi_B^{(q)}(0) - \varphi_A^{(q)}(0)\right), \tag{10b}$$

where

$$\varphi^{(m)}(\mathbf{r}) = \varphi(0, \mathbf{r}), \quad \varphi^{(d)}(\mathbf{r}) = \left[\frac{\partial \varphi(y, \mathbf{r})}{\partial y}\right]_{y=0}, \quad \varphi^{(q)}(\mathbf{r}) = \left[\frac{\partial^2 \varphi(y, \mathbf{r})}{\partial y^2}\right]_{y=0}, \tag{11}$$

$$\varphi^{(d)}(y) \equiv \varphi^{(d)}\left(\begin{pmatrix}0\\y\end{pmatrix}\right), \quad \varphi^{(q)}(y) \equiv \varphi^{(q)}\left(\begin{pmatrix}0\\y\end{pmatrix}\right).$$

The high frequency approximations to the longitudinal and the transverse wake functions of point charge read

$$w_\|(s, y_1, y_2) = cZ_\|(\mathbf{y})\delta(s), \tag{12}$$

$$w_y(s, y_1, y_2) = \theta(s)\omega Z_y(\omega, \mathbf{y}),$$

where $\theta(s)$ is the Heaviside step function and we have used the fact that in the optical approximation the product $\omega Z_y(\omega, \mathbf{y})$ is a constant independent from $\omega$.

The high frequency approximation of the loss factor for a bunch with arbitrary longitudinal profile $\lambda(s)$ can be written as

$$k_\|(y_1, y_2) = cZ_\|(\mathbf{y}) \int_{-\infty}^{-\infty} \lambda^2(s) ds.$$

For the Gaussian bunch with rms length $\sigma$ the loss factor reads

$$k_\|(y_1, y_2) = \frac{c}{2\sqrt{\pi}\sigma} Z_\|(\mathbf{y}). \tag{13}$$

The high frequency approximation of the transverse kick factor for an arbitrary normalized bunch shape $\lambda(s)$ can be found as

$$k_y(y_1, y_2) = \omega Z_y(\omega, \mathbf{y}) \int_{-\infty}^{\infty} \lambda(x) \int_{-\infty}^{x} \lambda(y) dy dx = \frac{\omega}{2} Z_y(\omega, \mathbf{y}), \tag{14}$$

where we have used the relation

$$\int_{-\infty}^{\infty} \lambda(x) \int_{-\infty}^{x} \lambda(y) dy dx = \frac{1}{2}\left(\int_{-\infty}^{\infty} \lambda(x) dx\right)^2.$$

Finally, for any structure the forward $Z_\|^+(\omega, \mathbf{r}_1, \mathbf{r}_2)$ and the backward $Z_\|^-(\omega, \mathbf{r}_1, \mathbf{r}_2)$ impedances in general case are related as [6, 7]

$$Z_\|^+(\omega, \mathbf{r}_1, \mathbf{r}_2) - Z_\|^-(\omega, \mathbf{r}_2, \mathbf{r}_1) = \frac{2}{Z_0 c^2}\left[\int_{S_B} \nabla\varphi_B(\mathbf{r}_1, \mathbf{r}) \nabla\varphi_B(\mathbf{r}_2, \mathbf{r}) ds - \int_{S_A} \nabla\varphi_A(\mathbf{r}_1, \mathbf{r}) \nabla\varphi_A(\mathbf{r}_2, \mathbf{r}) ds\right].$$

With the help of the Green's identity we can rewrite it as

$$Z_\|^+(\omega, \mathbf{r}_1, \mathbf{r}_2) - Z_\|^-(\omega, \mathbf{r}_2, \mathbf{r}_1) = \frac{2}{c}\left[\varphi_B(\mathbf{r}_1, \mathbf{r}_2) - \varphi_A(\mathbf{r}_1, \mathbf{r}_2)\right]. \tag{15}$$

## III. TRANSVERSE IMPEDANCE OF LASER MIRROR OF RF GUN

An electron beam with low emittance and high peak current is required in the European Free Electron Laser [1]. To meet these requirements, a laser-driven photocathode radio frequency (RF) gun has been designed and studied both experimentally and theoretically at DESY. The mirror in the vacuum chamber which reflects the laser beam on the cathode is a main source of the transverse kick on the axis. It could disturb an emittance compensation mechanism of the gun considerably. In this section we estimate numerically and analytically the transverse kick due to the mirror.

The geometry and the main dimensions of the vacuum mirror are shown in Fig. 2 ($d = 10$ mm, $a = 12.4$ mm, $R = 18.5$ mm). The mirror has a complicated three dimensional shape, which was used in numerical simulations. The bunch length at the position of the mirror is about of $\sigma = 2$ mm with a transverse size of about 2.6 mm. The parameters of the mirror and of the beam fulfil the relations $\sigma \ll a$, $L \ll a^2/\sigma$, where $L$ is the length of the mirror in z-direction. Hence, we can simplify the model to a thin iris and use the optical approximation to calculate the wake function.

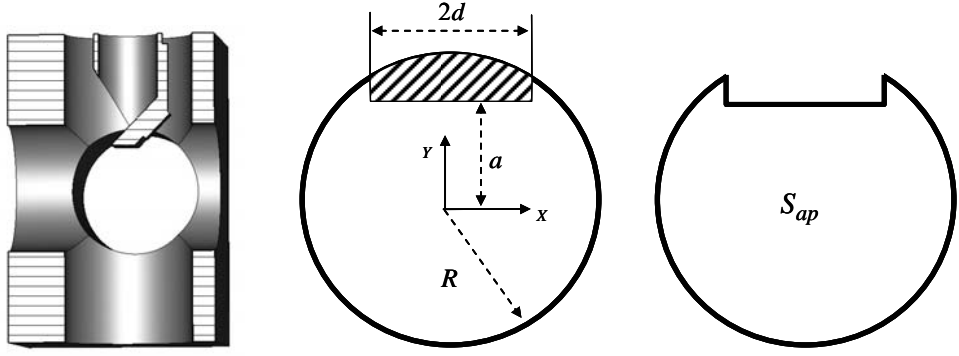

FIG 2. The vacuum mirror geometry and the aperture.

The Green's function for the Laplacian inside the circle of radius $R$ can be written as

$$\varphi_R(\mathbf{r}_1,\mathbf{r}) = \frac{1}{4\pi\varepsilon_0}\operatorname{Re}\left(2\ln\frac{z^*z_1 - R^2}{R(z-z_1)}\right) = \frac{1}{4\pi\varepsilon_0}\ln\frac{(x_1 y - xy_1)^2 + (xx_1 + yy_1 - R^2)^2}{R^2|\mathbf{r}-\mathbf{r}_1|^2},$$

$$\mathbf{r} = (x,y)^T, \quad z = x+iy.$$

For the studied geometry it holds that $\varphi_A = \varphi_B = \varphi_R$. The aperture $S_{ap}$ is shown in Fig. 2 on the right.

The monopole term of transverse impedance can be found from Eq. (8a)

$$Z_y^{(m)}(\omega) = -\frac{2\varepsilon_0}{\omega}\int_{\partial S_{ap}}\varphi_B^{(d)}(\mathbf{r})\partial_\mathbf{n}\varphi_A^{(m)}(\mathbf{r})dl =$$

$$= -\frac{2\varepsilon_0}{\omega}\left(2\int_0^d \frac{a^2(a^2+x^2-R^2)}{4\pi^2\varepsilon_0^2 R^2(a^2+x^2)^2}dx - 2\int_a^{\sqrt{R^2-d^2}}\frac{dy(d^2+y^2-R^2)}{4\pi^2\varepsilon_0^2 R^2(d^2+y^2)^2}dy\right) =$$

$$= \frac{1}{2\varepsilon_0\pi^2\omega aR^2}\left[(R^2-2a^2)\alpha + ad\left(1+\ln\frac{R^2}{a^2+d^2}\right)\right],$$

where we have used that $\varphi_B^{(d)}(\mathbf{r})\equiv 0$ for the boundary points $\mathbf{r}\in\partial S_B$.

The dipole term can be found from Eq. (9a)

$$Z_y^{(d)}(\omega) = -\frac{2\varepsilon_0}{\omega}\int_{\partial S_{ap}}\varphi_B^{(d)}(\mathbf{r})\partial_\mathbf{n}\varphi_A^{(d)}(\mathbf{r})dl =$$

$$= A^{-1}\left[aR^4d - 4a^3d^3 + R^4d^2\alpha - a^2\left(2d^3Q + R^4\beta + R^2d(Q-4d(\alpha+\beta))\right)\right].$$

Finally the quadrupole term reads (see Eq. (10a))

$$Z_y^{(q)}(\omega) = -\frac{2\varepsilon_0}{\omega}\int_{\partial S_{ap}}\varphi_B^{(m)}(\mathbf{r})\partial_\mathbf{n}\varphi_A^{(q)}(\mathbf{r})dl =$$

$$= \frac{1}{AB}\left[ad\left(R^4(d^2-a^2)+\left(a^2+d^2\right)\left(R^2+6d^2\right)aQ\right)+B\left(a^2\left(R^4-8d^4\right)\beta+\left(R^4-8a^4\right)d^2\alpha\right)\right],$$

$$\alpha = \tan^{-1}\left(\frac{d}{a}\right), \beta = \cot^{-1}\left(\frac{d}{Q}\right) - \tan^{-1}\left(\frac{a}{d}\right), Q = \sqrt{R^2-d^2}, B = a^2+d^2, A = 4\pi^2\varepsilon_0\omega a^2 R^4 d^2.$$

In order to check the above analytical results we have calculated the kick of the thin iris for a bunch with an RMS length $\sigma = 0.5$ mm. The results are shown in Table I. The analytical results do not depend on the bunch length.

TABLE I. Kick for iris and $\sigma = 0.5$ mm

|  | $k_y(0,0)$, V/pC | $k_y^{(d)}$, V/pC/m | $k_y^{(q)}$, V/pC/m |
|---|---|---|---|
| Analytical | 0.124 | 13.1 | 12.1 |
| Numerical | 0.120 | 13.1 | 11.6 |

Table 2 presents the numerical results from 3D calculation for the real mirror geometry (see Fig.2, the left picture) as obtained with code ECHO [4]. We see that the simple analytical model overestimates the kick on the axis $k_y(0,0)$, but underestimates the dipole term $k_y^{(d)}$.

TABLE II. Kick for mirror and $\sigma = 2$ mm

|  | $k_y(0,0)$, V/pC | $k_y^{(d)}$, V/pC/m | $k_y^{(q)}$, V/pC/m |
|---|---|---|---|
| Analytical | 0.12 | 13 | 12 |
| Numerical | 0.08 | 24 | 7.5 |

From ASTRA simulations [8] for the bunch with charge $Q = 1$ nC we have found a bunch energy of $E = 6.6$ MeV, an optical beta function of $\beta = 8.4$ m and an emittance of $\gamma\varepsilon_{0y} = 2.156$ μm at position of the mirror. Hence, as shown in Appendix, for the kick factor $k_y(0,0) = 0.124$ kV/nC the analytical estimation of the emittance growth gives

$$\frac{\varepsilon_y - \varepsilon_{0y}}{\varepsilon_{0y}} = 0.3\%.$$

## IV. TRANSVERSE IMPEDANCE OF OTR SCREENS

Several OTR screens will be installed in XFEL beamline [1]. The length of the screens in z-direction is much shorter than catch up distance, $L \ll a^2/\sigma$, and we can use the optical approximation to estimate the coupling impedances. The geometry of the screens is shown in Fig. 3. The main dimensions of the OTR screens are $R = 20.25$ mm, $h = 15$ mm, $d = 4...12.5\,\text{mm}$, $a = 2...10\,\text{mm}$. For the studied geometry the Green's function for the Laplacian inside the circle has to be used: $\varphi_A = \varphi_B = \varphi_R$. The aperture $S_{ap}$ is shown in Fig.3 on the right side.

The monopole term of the transverse impedance can be found as follows

$$Z_y^{(m)}(\omega) = -\frac{2\varepsilon_0}{\omega}\left(\int_{-d}^{\sqrt{R^2-a^2}} \frac{a^2\left(a^2+x^2-R^2\right)}{4\pi^2\varepsilon_0^2 R^2\left(a^2+x^2\right)^2}dx - \int_{-d}^{\sqrt{R^2-h^2}} \frac{h^2\left(h^2+x^2-R^2\right)}{4\pi^2\varepsilon_0^2 R^2\left(h^2+x^2\right)^2}dx\right) +$$

$$+\frac{2\varepsilon_0}{\omega}\int_a^h \frac{dy\left(d^2+y^2-R^2\right)}{4\pi^2\varepsilon_0^2 R^2\left(d^2+y^2\right)^2}dy = \frac{1}{4\pi^2\varepsilon_0\omega a R^2 h}\left[F(a,h)-F(h,a)\right],$$

$$F(x,y) = \left(R^2-2x^2\right)y\left(\cot^{-1}\left(\frac{x}{\sqrt{R^2-x^2}}\right)+\tan^{-1}\left(\frac{d}{x}\right)\right)+ay\left(\sqrt{R^2-x^2}+d\ln\left(d^2+y^2\right)\right).$$

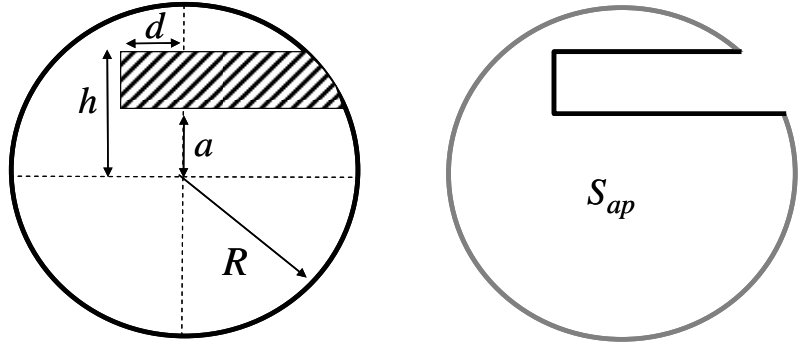

FIG. 3. The OTR geometry and its aperture.

The left graph in Fig. 4 shows the calculated transverse kick for different values of $d$ and $a$. It is independent from the bunch shape and the bunch length.

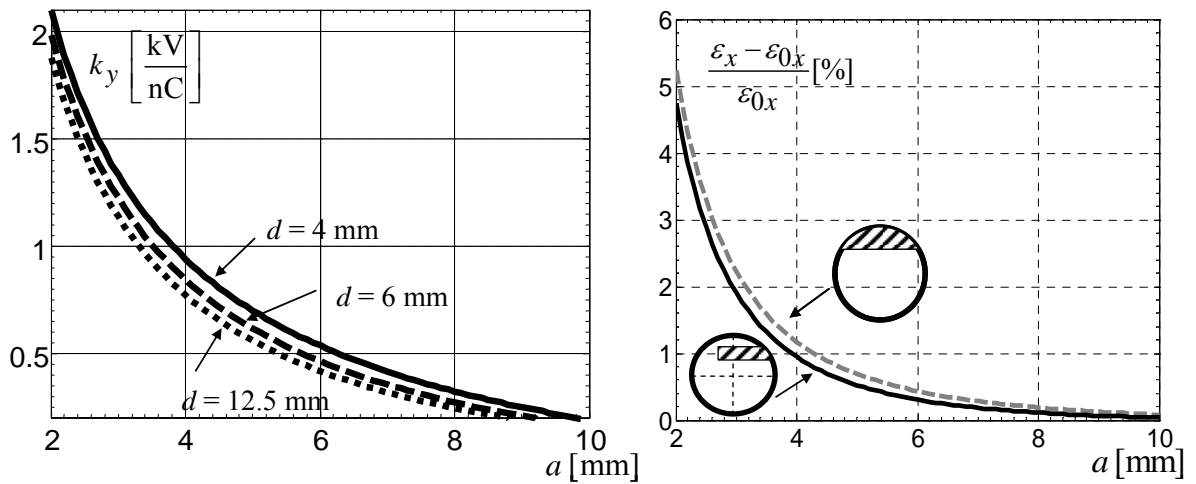

FIG. 4. The transverse kick and the emittance growth.

The right plot in Fig. 4 presents the calculated emittance growth for a OTR screen in the injector section ($d = 4\,\text{mm}$) for an arbitrary longitudinal bunch profile with emittance $\varepsilon_{0y} = 1\,\mu\text{m}$, energy $E = 130\,\text{MeV}$ and beta function $\beta = 4.3\,\text{m}$. The dashed line in the right plot shows the result for a simple geometric approximation by a circle segment

$$Z_y^{(m)}(\omega) = -\frac{2\varepsilon_0}{\omega}\left(2\int_0^{\sqrt{R^2-a^2}}\frac{a^2\left(a^2+x^2-R^2\right)}{4\pi^2\varepsilon_0^2 R^2\left(a^2+x^2\right)^2}dx\right) =$$

$$= \frac{1}{2\pi^2\varepsilon_0\omega a R^2}\left(\left(R^2-2a^2\right)\tan^{-1}\left(\frac{a}{\sqrt{R^2-a^2}}\right)+a\sqrt{R^2-a^2}\right).$$

## V. LONGITUDINAL IMPEDANCE OF ELLIPTICAL TO ROUND TRANSITIONS IN UNDULATOR SECTION

In the undulator intersection the vacuum chamber changes from an elliptical pipe to a round one. At the position of the elliptical-to-round transition (E2R) an elliptical absorber of a smaller cross-section is placed as shown in Fig. 5.

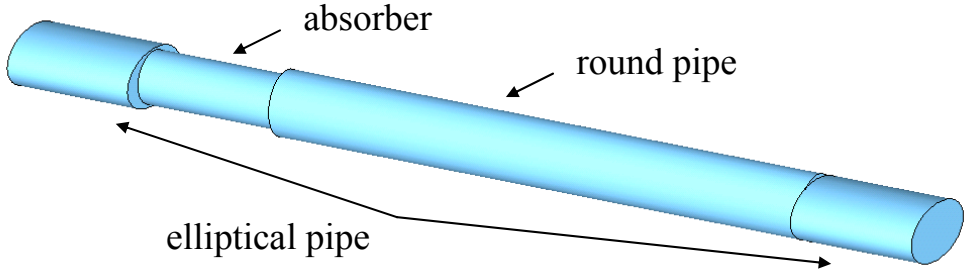

FIG. 5. The geometry of the vacuum chamber in the undulator intersections.

At the beginning let us consider a simple case without the absorber. In this case we have only an elliptical-to-round (E2R) pipe transition. The geometry of the transition is shown in Fig.6.

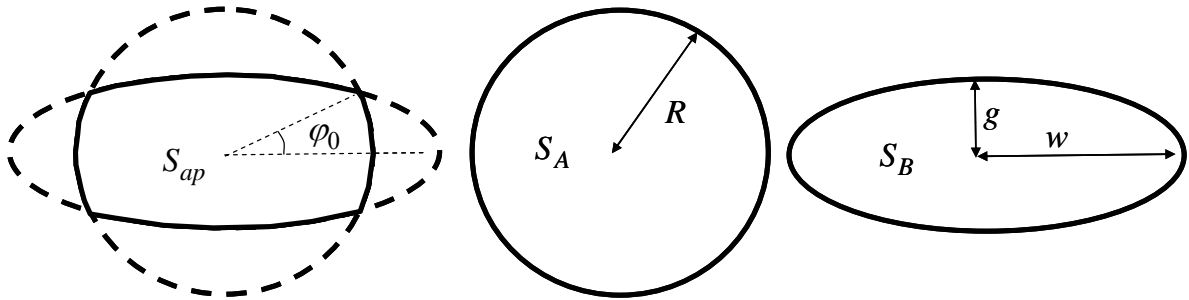

FIG 6. The geometry of transition from round pipe to elliptical one.

The Green's function for the Laplacian inside the ellipse with half axes $w$ and $g$ can be written as [9]

$$\varphi_E(\mathbf{r}_1,\mathbf{r}) = \varphi_E^0(\mathbf{r}_1,\mathbf{r}) - \varphi_E^0(\mathbf{r}_1,\mathbf{r}_0), \quad \mathbf{r}_0 = (0,g)^T,$$

$$\varphi_E^0(\mathbf{r}_1,\mathbf{r}) = -\frac{1}{\pi\varepsilon_0} \sum_{n=1}^{\infty} \frac{e^{-nu}}{n} \left[ \frac{\operatorname{Re}T_n\left(\frac{x+iy}{d}\right)\operatorname{Re}T_n\left(\frac{x_1+iy_1}{d}\right)}{\cosh(nu)} + \frac{\operatorname{Im}T_n\left(\frac{x+iy}{d}\right)\operatorname{Im}T_n\left(\frac{x_1+iy_1}{d}\right)}{\sinh(nu)} \right] -$$

$$-\frac{1}{4\pi\varepsilon_0}\ln\left[(x-x_1)^2 + (y-y_1)^2\right],$$

$$d = \sqrt{w^2 - g^2}, \quad u = \coth^{-1}\left(\frac{w}{g}\right),$$

where $T_n(z)$ are the Chebyshev polynomials of the first kind.

For the round-to-eliptical (R2E) pipe transition the Green's functions has to be assigned as $\varphi_A = \varphi_R$, $\varphi_B = \varphi_E$. The aperture $S_{ap}$ is shown in Fig. 6.

The longitudinal impedance on the axis can be found as

$$Z_{\|}^{(R2E)} = -\frac{2\varepsilon_0}{c} \int_{\partial S_{ap}} \varphi_E(\mathbf{0},\mathbf{r})\partial_\mathbf{n}\varphi_R(\mathbf{0},\mathbf{r})dl = -\frac{8\varepsilon_0}{c} \int_0^{\varphi_0} \varphi_E(\varphi,R)\partial_r\varphi_R(\varphi,R)Rd\varphi =$$

$$= \frac{4}{c\pi}\int_0^{\varphi_0} \varphi_E(\varphi,R)d\varphi, \tag{16}$$

where

$$\varphi_0 = \tan^{-1}\left(\frac{g}{w}\sqrt{\frac{w^2-R^2}{R^2-g^2}}\right), \quad \varphi_E(\varphi,r) \equiv \varphi_E(\mathbf{0},(r\cos\varphi,r\sin\varphi)^T),$$

$$\varphi_R(\varphi,r) \equiv \varphi_R(\mathbf{0},(r\cos\varphi,r\sin\varphi)^T),$$

and we have used the relation

$$\partial_r\varphi_R(\varphi,r) = -\frac{1}{2\pi\varepsilon_0 r}.$$

In order to calculate the longitudinal impedance of the elliptical-to-round (E2R) pipe transition we can use Eq. (15)

$$Z_{\|}^{E2R} = Z_{\|}^{R2E} - \frac{2}{c}\left[\varphi_E(\mathbf{0},\mathbf{0}) - \varphi_R(\mathbf{0},\mathbf{0})\right].$$

We evaluate the one dimensional integral (Eq.(16)) numerically. The right graph in Fig. 7 presents the results for the fixed size of the elliptical pipe ($w_0 = 0.75\,\text{cm}$, $g_0 = 0.44\,\text{cm}$) and the Gaussian beam with rms ength $\sigma = 25\,\mu\text{m}$. The black dots show the numerical results from CST Particle Studio obtained for the bunch length $\sigma = 100\,\mu\text{m}$ and scaled (see Eq. (13)) to the bunch length $\sigma = 25\,\mu\text{m}$.

Let us now consider the geometry with the absorber included. Here we consider the absorber as a long collimator. The absorber has the half width $w_1 = 0.45$ cm and the half height $g_1 = 0.4$ cm.

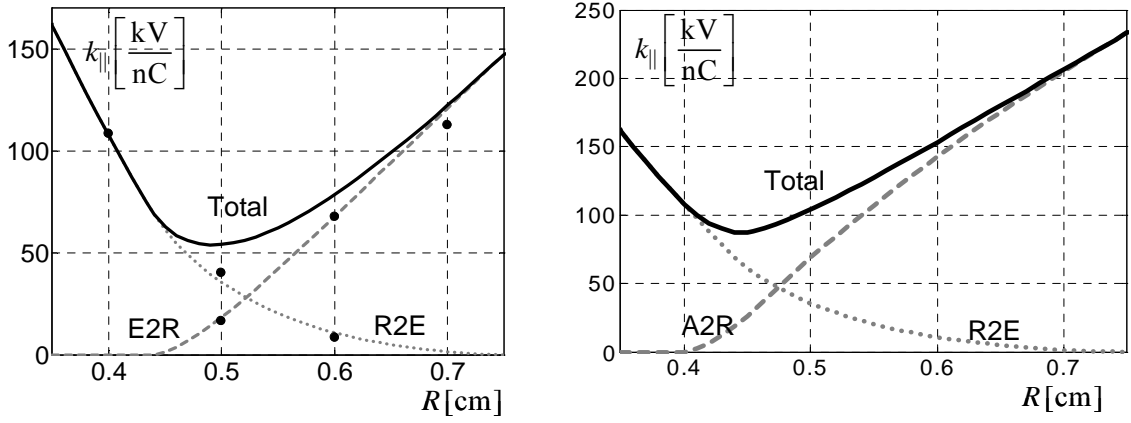

FIG 7. Dependence of the loss factor from the radius of the round pipe. The left graph presents the results without the absorber, the right graph presents the results with the absorber included. The black dots show the numerical results from CST Particle Studio.

The transition from the elliptical pipe to the absorber (E2A) can be considered as in-step transition and we have [3]
$$Z_\parallel^{(E2A)} = 0.$$

The contribution of the absorber to round pipe (A2R) transition can be found from Eq. (16) with $w = w_1$ and $g = g_1$. The final result is presented in Fig. 7 in the right graph. We can conclude that the optimal radius of the round pipe in the undulator intersection is 45-50 mm.

## VI. LONGITUDINAL IMPEDANCE OF ROUND TO RECTANGULAR TRANSITIONS IN BUNCH COMPRESSORS

The bunch compression scenarios for the XFEL have been laid out such, that a broad range of parameters can be obtained. To have a large range of variation in the deflecting angle in the bunch compressors a rectangular beam pipe will be used there. Below we consider the longitudinal wake due to round-to-rectangular (R2rct) pipe transition in the vacuum chamber. The geometry of the transition is shown in Fig. 8.

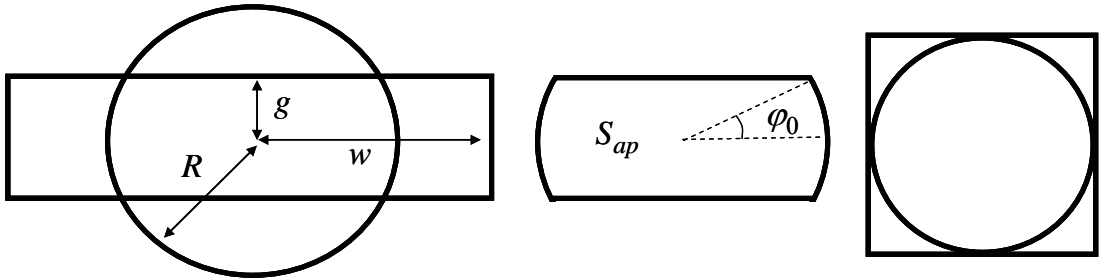

FIG. 8. The geometry of transition from round pipe to rectangular one.

The Green's function for the Laplacian inside of the rectangle of width $2w$ and height $2g$ for $\mathbf{r}_1 = (0, y_1)^T$ can be written as [9]

$$\varphi_{rct}(\mathbf{r}_1, \mathbf{r}) = -\frac{1}{\pi\varepsilon_0} \sum_{n=1}^{\infty} \frac{e^{-\frac{n\pi w}{2g}}}{n \cosh\frac{n\pi w}{2g}} \sin\frac{n\pi}{2g}(y+g) \sin\frac{n\pi}{2g}(y_1+g) +$$

$$+ \frac{1}{4\pi\varepsilon_0} \ln\left( \left[x^2 + (y-y_1)^2\right] \frac{\sinh^2\frac{\pi x}{4g} + \cos^2\frac{\pi}{4g}(y+y_1)}{\sinh^2\frac{\pi x}{4g} + \sin^2\frac{\pi}{4g}(y-y_1)} \right) - \frac{1}{4\pi\varepsilon_0} \ln\left[(x-x_1)^2 + (y-y_1)^2\right].$$

For the round-to-rectangular pipe transition the Green's function has to be chosen as $\varphi_A = \varphi_R$, $\varphi_B = \varphi_{rct}$. The aperture $S_{ap}$ is shown in Fig. 8 at the right side.

The longitudinal impedance on the axis can be found from Eq. (7)

$$Z_\parallel^{(R2rct)} = -\frac{2\varepsilon_0}{c} \int_{\partial S_{ap}} \varphi_{rct}(\mathbf{0},\mathbf{r}) \partial_\mathbf{n} \varphi_R(\mathbf{0},\mathbf{r}) dl = \frac{4}{c\pi} \int_0^{\varphi_0} \varphi_{rct}(\varphi, R) d\varphi,$$

where

$$\varphi_0 = \tan^{-1}\left(\frac{g}{\sqrt{R^2 - g^2}}\right), \quad \varphi_{rct}(\varphi, R) \equiv \varphi_{rct}(\mathbf{0}, (R\cos\varphi, R\sin\varphi)^T).$$

In order to calculate the rectangular to round (rct2R) transition we use Eq. (15)

$$Z_\parallel^{rct2R} = Z_\parallel^{R2rct} - \frac{2}{c}\left[\varphi_{rct}(\mathbf{0},\mathbf{0}) - \varphi_R(\mathbf{0},\mathbf{0})\right].$$

We calculate the above expressions numerically. The left plot in Fig. 9 presents the results for the fixed size of the rectangular pipe ($w = 10\,\text{cm}$, $g = 4\,\text{cm}$). The right plot in Fig. 9 presents the results for the fixed size of the round pipe ($R = 5\,\text{cm}$, $w = 10\,\text{cm}$).

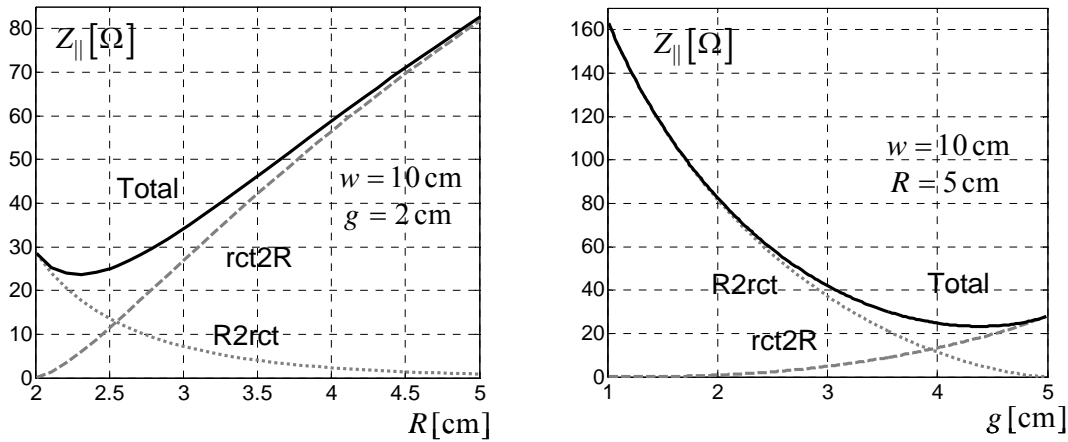

FIG. 9. The longitudinal impedance in the bunch compessor.

Finally, we have calculated a special case of the round-to-square (R2S) pipe transition shown in Fig.8 at the right side. The impedance of such transition is independent from the geometrical parameters

$$Z_{\|}^{(R2S)} = 9\,\Omega,\ Z_{\|}^{(S2R)} = 0.$$

## VII. LONGITUDINAL AND TRANSVERSE IMPEDACES OF ROUND MISALIGNED PIPES

The XFEL beam line has a length of several kilometres. It consists of many short beam pipes connected by bellows. In this section we estimate the effect of the round pipe misalignment.

Let us consider only one pipe shifted from the axis by distance $2g$. The geometry of the transition is shown in Fig. 10 and Fig. 11.

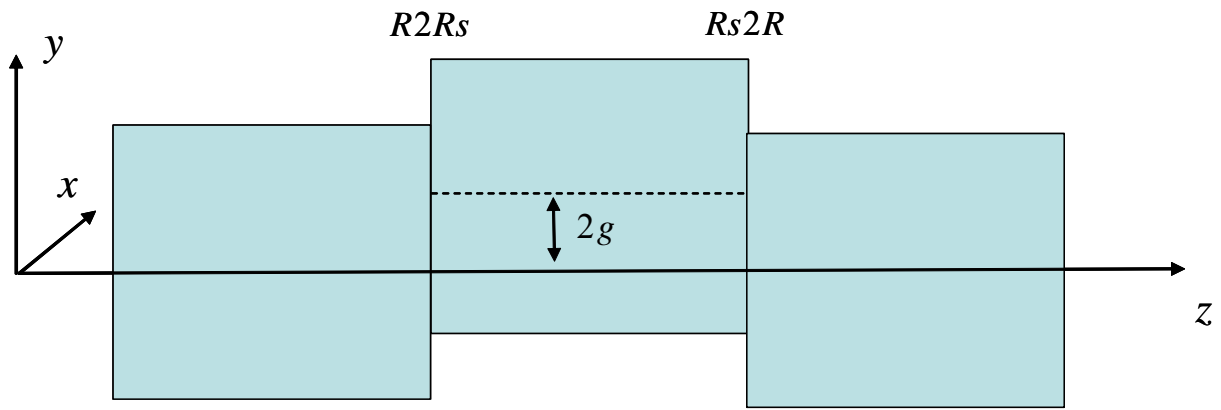

FIG. 10. The geometry of the misaligned pipes

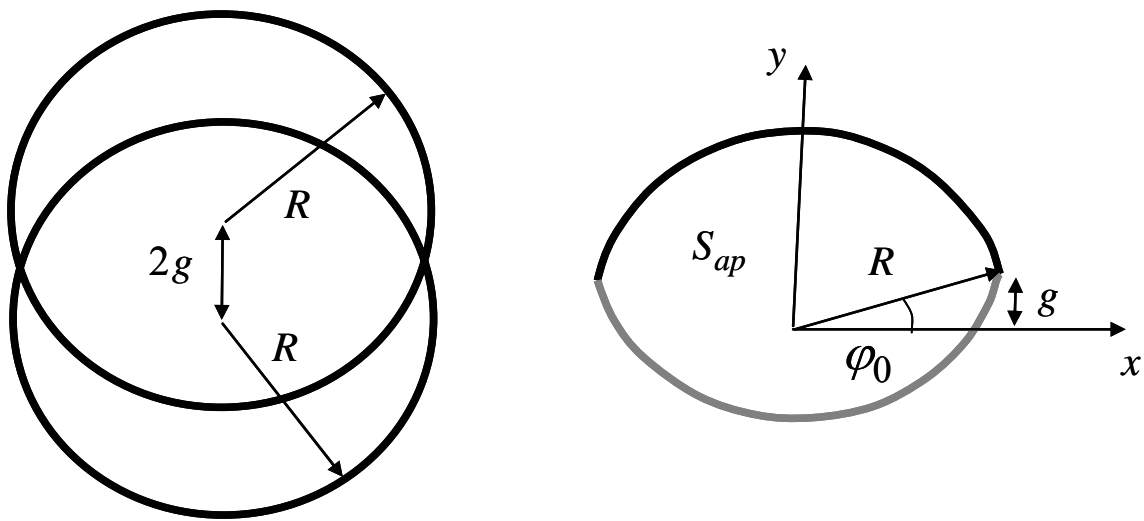

FIG. 11. The aperture of the misaligned pipes and the integration path.

For the round pipe to the shifted round pipe (R2Rs) transition the Green functions can be written as

$$\varphi_A(\mathbf{r}_1,\mathbf{r}) = \varphi_R(\mathbf{r}_1,\mathbf{r}), \quad \varphi_B(\mathbf{r}_1,\mathbf{r}) = \varphi_R(\mathbf{r}_1 - 2\mathbf{r}_0, \mathbf{r} - 2\mathbf{r}_0), \quad \mathbf{r}_0 = (0,g)^T.$$

The aperture $S_{ap}$ is shown in Fig. 11. We consider two positions of the beam: on the axis ($y=0$) and in the middle between the pipes ($y=g$).

The longitudinal impedance at the axis can be found as

$$Z_\parallel^{(R2Rs)}(0) \equiv Z_\parallel^{(R2Rs)}(0,0) = -\frac{2\varepsilon_0}{c} \int_{\partial S_{ap}} \varphi_R(-2\mathbf{r}_0, \mathbf{r} - 2\mathbf{r}_0) \partial_\mathbf{n} \varphi_R(\mathbf{0},\mathbf{r}) dl =$$

$$= \frac{Z_0}{2\pi^2} \int_0^{\varphi_0} \ln\left(16\alpha^4 + (1-4\alpha^2)(1+4\alpha \sin\varphi)\right) d\varphi,$$

where

$$\alpha = \frac{g}{R}, \quad \varphi_0 = \sin^{-1}\left(\frac{g}{R}\right).$$

The longitudinal impedance for the beam moving in the middle between the pipes at the position $\mathbf{r}_0 = (0,g)^T$ can be found as

$$Z_\parallel^{(R2Rs)}(g) \equiv Z_\parallel^{(R2Rs)}(g,g) = -\frac{2\varepsilon_0}{c} \int_{\partial S_{ap}} \varphi_R(-\mathbf{r}_0, \mathbf{r} - 2\mathbf{r}_0) \partial_\mathbf{n} \varphi_R(\mathbf{r}_0,\mathbf{r}) dl =$$

$$= \frac{Z_0}{2\pi^2} \int_0^{\varphi_0} \frac{(1-\alpha^2)}{1+\alpha^2 - 2\alpha \sin\varphi} \ln\left[2\alpha^2 - 1 + \frac{2(\alpha^2-1)^2}{1+\alpha^2 - 2\alpha \sin\varphi}\right] d\varphi,$$

where $Z_0 = (c\varepsilon_0)^{-1}$ is an impedance of the vacuum.

The both impedances are shown in Fig. 12. The numerical results obtained by direct numerical solution of the Maxwell's equations are shown by black points at these plots (and the plots in Fig. 13) as well. The numerical simulations are done with CST Particle Studio for the round pipe with radius $R=3$ mm and the Gaussian bunch with rms length $\sigma = 0.1$ mm.

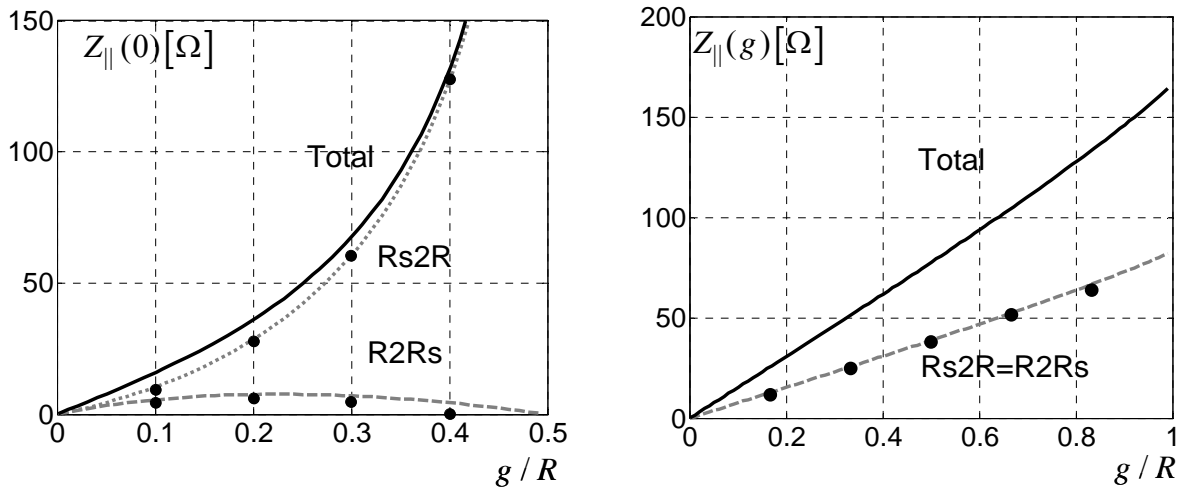

FIG. 12. The longitudinal impedance of the misalighned pipes. The black dots show the numerical results from CST Particle Studio.

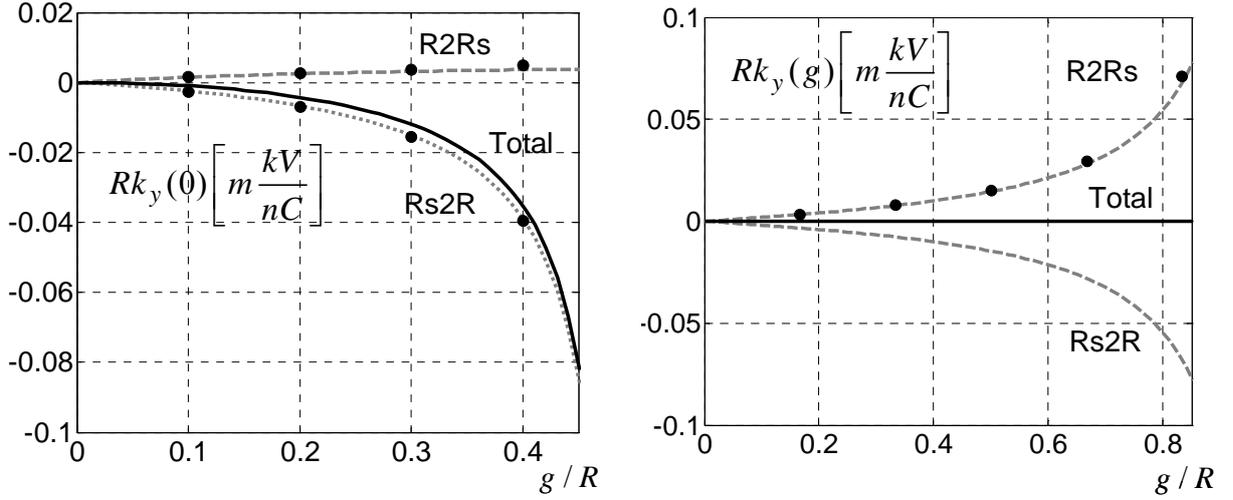

FIG. 13. The transverse kick of the misaligned pipes. The black dots show the numerical results from CST Particle Studio.

The transverse kick at the axis can be found in closed analytical form and reads

$$k_y^{(R2Rs)}(0) \equiv k_y^{(R2Rs)}(0,0) = -\frac{Z_0 c}{8\pi^2}\left(F_0\left(\frac{\pi}{2}\right) - F_0(\varphi_0)\right), \quad (17)$$

$$F_0(\varphi) = \frac{4\cos\varphi}{R} + \frac{8\varphi\alpha^2 - \varphi + 2\tan^{-1}[x,y]}{R(4\alpha^3 - \alpha)},$$

$$x = (4\alpha^2 - 1)\cos\frac{\varphi}{2} - 2\alpha\sin\frac{\varphi}{2}, \quad x = (4\alpha^2 - 1)\sin\frac{\varphi}{2} - 2\alpha\cos\frac{\varphi}{2},$$

$$k_y^{(Rs2R)}(0) \equiv k_y^{(Rs2R)}(0,0) = \frac{Z_0 c}{8\pi^2}\left(F_{2g}\left(\frac{\pi}{2}\right) - F_{2g}(\varphi_0)\right), \quad (18)$$

$$F_{2g}(\varphi) = \frac{4\alpha^2 - 1}{R\alpha}\left(\varphi - \frac{2(1+16\alpha^2(\alpha^2-1))\tan^{-1}[x_2, y_2]}{(1-4\alpha^2)^2} + \frac{4\alpha\cos\varphi}{16\alpha^4 - 1 + (4\alpha - 16\alpha^3)\sin\varphi}\right),$$

$$x_2 = 2\alpha\cos\frac{\varphi}{2} - \sin\frac{\varphi}{2}, \quad y_2 = \cos\frac{\varphi}{2} - 2\alpha\sin\frac{\varphi}{2}.$$

Expression $\tan^{-1}[x, y]$ defines an inverse tangent function $\arctan(y/x)$ taking into account which quadrant the point $(x, y)$ is in.

The transverse kick for the bunch moving between the pipes at the position $\mathbf{r}_0 = (0, g)^T$ can be found as

$$k_y^{(R2R)}(g) \equiv k_y^{(R2R)}(g,g) = -\frac{Z_0 c}{8\pi^2}\left(F_g\left(\frac{\pi}{2}\right) - F_g(\varphi_0)\right),$$

$$F_g(\varphi) = -\frac{2}{R\alpha(\alpha^2-1)}\left((1-8\alpha^2)\tan^{-1}(x_0, y_0) - \tan^{-1}(x_1, y_1) + 2\alpha(\alpha^2-1)\left(\frac{1}{1+\alpha^2} - \frac{\cos\varphi}{1+\alpha^2 - 2\alpha\sin\varphi}\right)\right),$$

$$x_0 = \alpha\cos\frac{\varphi}{2} - \sin\frac{\varphi}{2}, \quad y_0 = \cos\frac{\varphi}{2} - \alpha\sin\frac{\varphi}{2},$$

$$x_1 = (2\alpha^2 - 1)\cos\frac{\varphi}{2} - \alpha\sin\frac{\varphi}{2}, \quad y_1 = (2\alpha^2 - 1)\sin\frac{\varphi}{2} - \alpha\cos\frac{\varphi}{2},$$

$$k_y^{(Rs2R)}(g,g) = -k_y^{(R2Rs)}(g,g).$$

In the current design of XFEL [1] there are two bellows in each undulator intersection. In each bellow we will have a gap of about 2…5 mm between the pipes. Let us compare the effect of the gap to the effect of the pipe misalignment.

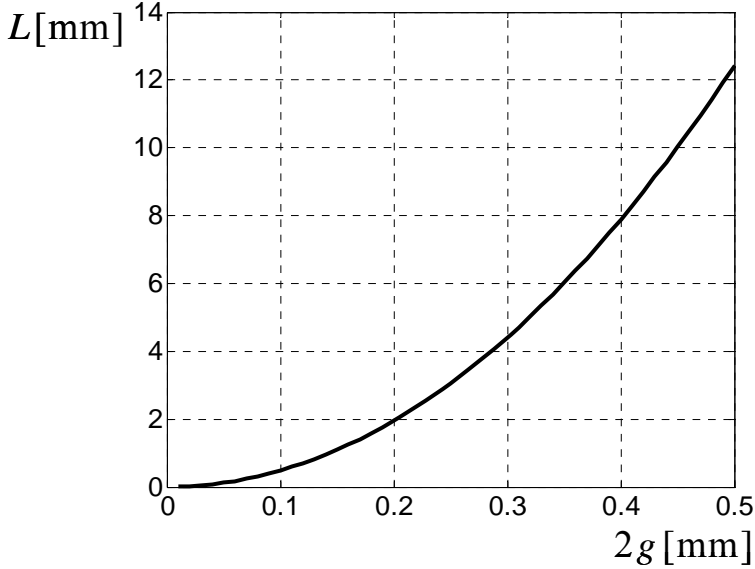

FIG. 14. The gap length L which produces the same energy loss as the round pipe shifted by 2g for the Gaussian bunch with rms length $\sigma = 25\,\mu m$.

A gap of length $L$ between round pipes of radius $R$ can be considered as a short pillbox cavity. The optical theory predicts zero impedance for a short cavity ($L \ll R^2/\sigma$). A long cavity can be treated in the optical approximation as a combination of a step-out and step-in transitions.

For the situation between these two extremes the diffraction model [10] could be applied. Then the longitudinal and the transverse wake functions, the loss and kick factors (for the Gaussian bunch with rms length $\sigma$) can be written as [10]

$$w_\parallel(s) = \theta(s)\frac{Z_0 c}{\sqrt{2}\pi^2 R}\sqrt{\frac{L}{s}}, \qquad k_\parallel(0) = \frac{Z_0 c}{4R\pi^{2.5}}\Gamma\left(\frac{1}{4}\right)\sqrt{\frac{L}{\sigma}},$$

$$w_r(s,r) = r\theta(s)\frac{2}{R^2}\frac{\sqrt{2}Z_0 c}{\pi^2 R}\sqrt{Ls}, \qquad k_\perp(r) = r\frac{2}{R^3}\frac{Z_0 c}{\pi^{2.5}}\Gamma\left(\frac{3}{4}\right)\sqrt{L\sigma}.$$

For the pipe of radius $R = 5\,\text{mm}$ with the gap length $L = 5\,\text{mm}$ and the bunch length of $\sigma = 25$ μm we obtain the energy loss of the pipe gap $k_\parallel(0) = 16.6\,\text{kV/nC}$. Fig. 14 presents the equivalent gap length $L$ between the round pipes which produces the same energy loss as the round pipe shifted by distance $2g$ from the axis. For the beam moving on the axis the energy loss of $16.6\,\text{kV/nC}$ will be obtained for the pipe shifted by $2g = 0.32\,\text{mm}$ from the axis.

For the same parameters and the beam moving at the distance $r = 1\,\text{mm}$ from the axis the transverse kick of the pipe gap is $k_\perp(r = 1\,\text{mm}) = 0.05\,\text{kV/nC}$. It follows from Eq. (17) and Eq. (18) for the beam moving on the axis that such kick will be obtained from the misaligned pipe with the shift parameter $\alpha = gR^{-1} = 0.052$.

## VIII. CONCLUSION

In this report we have applied the method of the optical approximation to estimate the high frequency impedances of different transitions in the vacuum chamber of the European XFEL. The bunch used for XFEL operation is very short and the analytical results obtained in this report are very good approximations to the coupling impedances. Almost all analytical results presented are new and supplement those already presented in [3]. The method of the optical approximation [2] is powerful and allows to study analytically a truly large class of transitions.

## ACKJNOWLEDGMENTS

We thank R. Wanzenberg for useful discussions and corrections.

## APPENDIX: ANALYTICAL ESTIMATION OF EMITTANCE GROWTH

According to the conditions of the optical approximation the wake potential of an arbitrary bunch profile $\lambda(s)$ can be found as

$$W_y(s) = 2k_y \Lambda(s), \quad \Lambda(s) = \int_{-\infty}^{s} \lambda(s')ds'.$$

The kick due to the bunch can be found as

$$\Delta y'(s) = \frac{\Delta p_y}{p_z} = \frac{eQW_y(s)}{\beta_z^2 E} = 2S\Lambda(s), \quad S = \frac{eQk_y}{\beta_z^2 E}.$$

Let us consider a transverse Gaussian particle distribution

$$\rho_0(y, y', s) = \frac{1}{2\pi\varepsilon_{0y}} \exp\left(-\frac{\gamma y^2 + 2\alpha y y' + \beta y'^2}{2\varepsilon_{0y}}\right) \lambda(s)$$

with an arbitrary longitudinal profile $\lambda(s)$.
After the kick the distribution has the form

$$\rho = \frac{\lambda(s)}{2\pi\varepsilon_{0y}} \exp\left(-\frac{\gamma y^2 + 2\alpha y(y' + \Delta y'(s)) + \beta(y' + \Delta y'(s))^2}{2\varepsilon_{0y}}\right).$$

The projected distribution after the kick does not depend on $\lambda(s)$

$$\bar{\rho} = \int_{-\infty}^{\infty} \rho(y, y', s)ds = \int_{0}^{1} \frac{1}{2\pi\varepsilon_{0y}} e^{-\frac{\gamma y^2 + 2\alpha y(y' + 2S\Lambda) + \beta(y' + 2S\Lambda)^2}{2\varepsilon_{0y}}} d\Lambda =$$

$$= \frac{Erf\left(\frac{y\alpha + (2S + y')\beta}{\sigma_{0y}\sqrt{2}}\right) - Erf\left(\frac{y\alpha + y'\beta}{\sigma_{0y}\sqrt{2}}\right)}{4\sqrt{2}\pi S \sigma_{0y}} e^{-\frac{y^2}{2\sigma_{0y}^2}},$$

where $\sigma_{0y} = \sqrt{\varepsilon_{0y}\beta}$.

The projected emittance can be calculated analytically

$$\varepsilon_y = \sqrt{\langle y^2 \rangle \langle y'^2 \rangle - \langle yy' \rangle} = \sqrt{\varepsilon_{0y}^2 + S^2 \frac{\varepsilon_{0y}\beta}{3}} \approx \varepsilon_{0y} + S^2 \frac{\beta}{6}$$

Hence, the relative emittance growth reads

$$\frac{\varepsilon_y - \varepsilon_{0y}}{\varepsilon_{0y}} = \sqrt{1 + S^2 \frac{\beta}{3\varepsilon_{0y}}} - 1 \approx S^2 \frac{\beta}{6\varepsilon_{0y}}.$$